\documentclass[prb,preprint]{revtex4}
\usepackage{amsfonts}
\usepackage{amsmath}
\usepackage{graphicx,subfigure}
\usepackage{epsfig}
\usepackage{dcolumn}
\usepackage{multirow}
\usepackage{bm}
\begin{document}
 \title{collapse of the magnetic moment under pressure of AFe$_2$ (A = Y, Zr, Lu and Hf) in the cubic Laves phase}
 \author{Wenxu Zhang and Wanli Zhang}
 \affiliation{State Key Laboratory of Electronic Thin Films and Integrated Devices,
 University of Electronic Science and Technology of China, Chengdu, 610054, P. R. China}
 \date{\today}
\begin{abstract}
The electronic structures of four Laves phase iron compounds (e.g. YFe$_2$, ZrFe$_2$, LuFe$_2$ and HfFe$_2$) have been 
calculated by the state-of-the-art full potential electronic structure code. The magnetic moments collapse under
hydrostatic pressure. This feature is found to be universal in these materials. Its electronic origin is provided by 
the sharp peaks in the density of states near the Fermi level. It is shown that a first order quantum phase transition
can be expected under pressure in Y(Zr, or Lu)Fe$_2$, while a second order one in HfFe$_2$. The bonding characteristics 
are discussed to elucidate the equilibrium lattice constant variation. The large spontaneous volume magnetostriction
gives one of the most important character of these compounds. Invar anomalies in these compounds
can be partly explained by the current work when the fast continuous magnetic moment decrease 
at the decrease of the lattice constant was properly considered. This work may remind the experimentalists of these ``old''
compounds and exploration of the quantum properties under high pressures are greatly encouraged.  
\end{abstract}
\maketitle
\section{Introduction}
Magneto-structural interaction is a fundamental phenomenon in magnetic functional materials.
The martensitic transformation caused by magnetization in magnetic shape memory is a direct magnification of this  
effect.\cite{kiefer} Because of the huge variation of the dimensions, the compounds can be used in sensors and actuators.
Meanwhile, this effect can also compensate the variation of the volume due to temperatures, as shown in the so-called Invar effect\cite{wasserman}. 
It was proposed to be connected with magnetism because the plateau of the volume expansion starts apparently almost at the
Curie temperature below which ordered magnetic moments are established. The nearly zero volume expansion 
is because that the Invar alloy has a spontaneous volume magnetostriction large enough to compensate normal thermal expansions 
due to lattice vibrations as suggested in the so-called magnetostriction model of Invar\cite{khmeloevskyi}, one of the many models to explain this
phenomenon.
\par The cubic Laves phase iron compounds show the Invar effect in stoichiometry which may serves as a simpler model system
for developing theory of Invar avoiding disorder complexing. It excludes the noncollinear magnetism and antiferromagnetic
moment as the mechanism of Invar anomaly\cite{khmeloevskyi}. ZrX$_2$ can be used as a hydrogen
storage materials \cite{hong} because of its suitable binding energy at the interstices. It was found that H-induced lattice expansion
will cause an increase in the magnetic moment.
They were found interesting even half century ago. As suggested in the early work
of Klein \emph{et al.}，\cite{klein} the electron-phonon coupling in the hypothetical paramagnetic ZrFe$_2$ is substantial
and the superconductivity transition temperature can be as high as 9 K.
\par Early work in AFe$_2$ found that A and Fe atomcs are in ferrimagnetic order.
Asano\cite{asano} studied the phase stability by comparing the total energies of different
phases (nonmagnetic, ferromagnetic, and antiferromagnetic states of C14 or C15 Laves phases).
He concludes that Y, Zr and Hf compounds are ferromagnetic C15 Laves phase at the ground state,
which is in agreement with the experiments.
Related properties of Laves phase iron compounds (e.g. YFe$_2$, ZrFe$_2$, LuFe$_2$ and HfFe$_2$) were studied in the past. 
Yamada\cite{yamada} has calculated the high
field susceptibility $\chi_{hf}$ of ZrFe$_2$ being $5.8\times10^{-4}$ emu/mol, and YFe$_2$
being $5.57\times10^{-4}$ emu/mol, which agree with the experimental
values $6.1\times10^{-4}$ emu/mol and $1.55\times10^{-4}$ emu/mol reasonably well.
Wortmann\cite{wortmann} showed that the hyperfine field decreases to zero at about 40 GPa in LuFe$_2$, and 50 GPa in 
YFe$_2$ at room temperature by nuclear forward scattering. At low temperature, the loss of magnetism took place at about 
90 GPa. Direct information of the magnetic moment under pressures was reported by Armitage\cite{armitage}. 
The measured $\frac{\partial ln \sigma}{\partial p}$'s are $-8.2\pm0.4$ and $-6.3\pm0.3\times10^{-4}$ kbar$^{-1}$ 
for YFe$_2$ and ZrFe$_2$, respectively. The experiments reported lattice constants, saturation
magnetization and Curie temperatures are listed in Table \ref{tab:exp}.
\begin{table}
\caption{The experimental (exp.) and calculated values by LDA and GGA of the lattice constant (a$_0$), spontaneous volume magnetostriction($\omega_s$) and magnetic moment (M$_s$) of AFe$_2$(A = Y, Zr, Hf and Lu) compounds. M$_s$ was obtained at T$=4.2$ K.}
\label{tab:exp}
\begin{tabular}{cccccccccccc}
\hline
\multirow{2}{*}{AFe$_2$} &\multicolumn{3}{c}{a({\AA})} &&\multicolumn{3}{c}{$\omega_s$($\times 10^{3}$)}&&\multicolumn{3}{c}{M$_s$($\mu_B$)}\\
\cline{2-4}\cline{6-8}\cline{10-12}
&exp.&LDA&GGA&&exp.&LDA&GGA&&exp.&LDA&GGA\\
\hline
YFe$_2$     & 7.363 &7.04 &7.28 & &small&49&55 & &2.90 &2.57&3.21 \\
ZrFe$_2$    & 7.06  &6.84 &7.04 & &10   &15&54 & &3.14 &2.38&3.10 \\
HfFe$_2$    & 7.02  &6.82 &7.00 & &8    &35&48 & &3.36 &2.86&3.26 \\
LuFe$_2$    & 7.217 &6.93 &7.15 & &$\sim$ &25&47&&2.97 &2.53&3.06 \\
\hline
\end{tabular}
\end{table}

\par Density functional theory (DFT) was a powerful tool to explain and predict the magnetic moment under pressure. 
For example, the HS-LS transition of transition metal monoxides (e.g. FeO, MnO, etc.) under the hydrostatic pressure 
as high as about 200 GPa were predicted by Cohen\cite{cohen}. Magnetic transition in these highly correlated insulators is 
the results of competition among the kinetic energy, exchange energy and Coulombic repulsion\cite{wxz}. 
The magnetic collapse in metals on the other hand can be qualitatively understood with the help of the Stoner model: 
In a simplified version of this model, the magnetic state is stable if $IN(E_F)>1$, where $I$ is the Stoner parameter, 
which is weakly dependent on the atomic distance, while  the density of states at the Fermi level $N(E_F)$ decreases as 
the band width increases under the pressure. At a certain critical pressure, the criterion is no longer satisfied, then 
the ferromagnetism cannot be sustainable. 

\par In this work, we found that there may exist a first order quantum phase transition under pressure in these compound, 
which is similar to the well studied case of ZrZn$_2$, MnSi, etc. Large volume magnetostriction, which is an 
Invar character of these compounds was presented. We further suggested that in order to show Invar effect, the magnetic moment
decrease with the volume should scale with that with the temperature.
\section{calculation details}
The C15 structure Laves phase (space group F$d\bar{3}m$)
has two formula units per face centered cubic unit cell. The full-potential local orbital minimum
basis band structure code (FPLO)\cite{fplo} was used in our calculation. Both the local spin density
approximation (LSDA)\cite{lsda} and general gradient approximation (GGA)\cite{pbe} of the exchange correlations functionals
were used here and the resuts were compared when necessary. The number
of $k$-points in the full Brillouin zone (BZ) is $30\times 30\times 30$, which can guarantee the
convergence of the total energy to microHartree. The scalar relativistic treatment was used where all
the relativistic effects were included except the spin-orbital coupling. The fixed spin moment (FSM)
calculations were used to investigate the possible multiple local energy minima with respect to the
magnetic moment. The spontanous volume-magnetostriction $\omega_s$ is defined in terms of the ratio of the
equilibrium volumes in the ferromagnetic FM (V$_{FM}$) and the paramagnetic PM state (V$_{PM}$)
\begin{equation}
\omega_s=\frac{V_{FM}-V_{PM}}{V_{PM}}.\label{eq:omegas}
\end{equation}
\section{the ground state properties}
The calculated lattice constant($a$), the total magnetic moment at equilibrium are
listed in Table \ref{tab:exp}. The agreement with the experiments is reasonably good. The lattice constant from LSDA is
lower than the experimental ones, which is notorious. The GGA results show a much better agreement. However,
the volume magnetostriction is largely overestimated by GGA than LSDA.
The systematic tendency is that the lattice constant and magnetic moments of Y and Lu compounds are smaller than these
of Zr and Hf compounds, where one more $d$ electron is added in the latter compounds. 
\par
\begin{figure}
\centering
\includegraphics[width=0.8\textwidth]{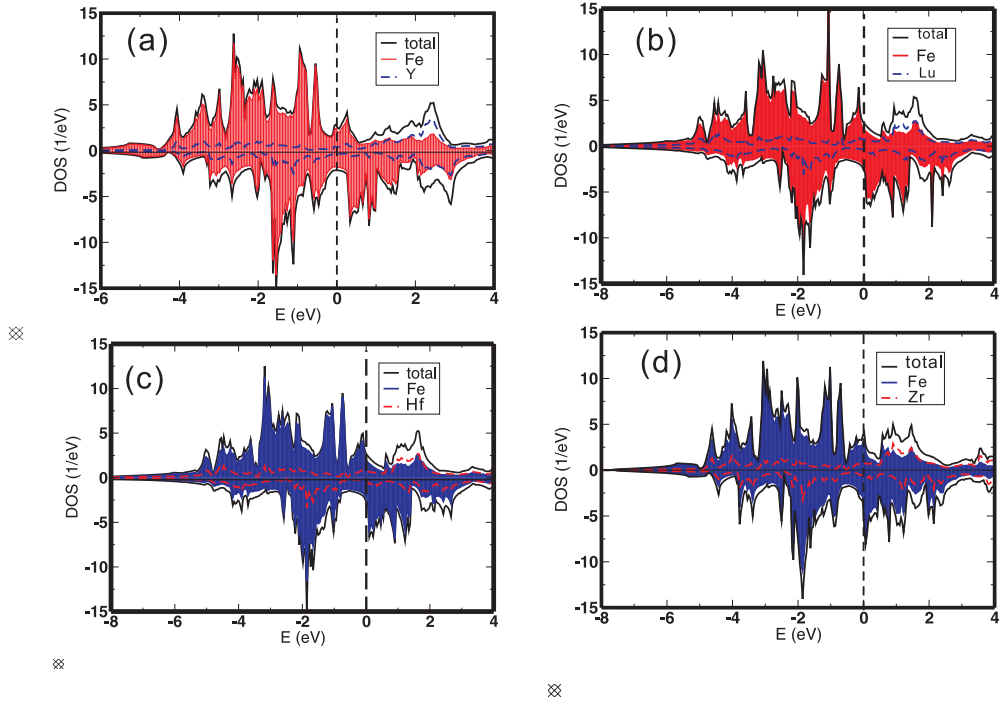}
\caption{\label{fig:dos-eq}(Color online) Density of states (DOS) and the atomic projection (shaded area) of
AFe$_2$(A=Y, Zr, Hf and Lu) at the ground state.}
\end{figure}

  The DOS at the equilibrium lattice constant of the selected compounds are shown in Figure {\ref{fig:dos-eq}}.
Common features of these DOS are quite obvious. The total DOS are mainly contributed by Fe's, the other elements only
show noticeable contributions at energy about 0.5 eV above the Fermi level. The bandwidth of the 3d states from Fe is about 
5 eV, which is a typical value of itinerant system. These states are responsible for the properties of these compounds.
Above the Fermi energy, there is relatively high DOSs in the minority spin channel. This feature is dominated by the anti-bonding 3$d$ states of Fe. The
bonding and antibonding states of the minority are separated by a deep and wide valley with
width about 1 eV near the Fermi level.
\begin{figure}[!htbp]
\centering
\includegraphics[width=0.8\textwidth]{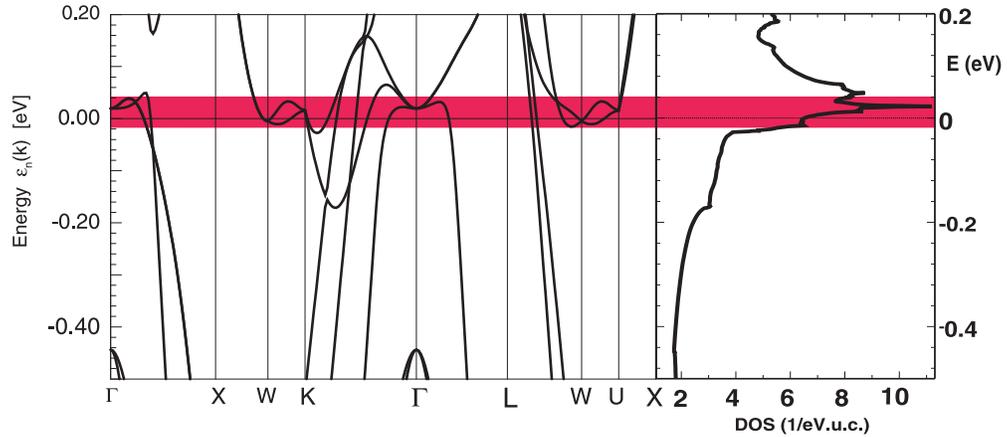}
\caption{The band structure and
the DOS of the minority spin channel of ZrFe$_2$ at the theoretical equilibrium lattice constant. The
rectangular block with height of 0.06 eV highlights the energy window around the pronounced DOS.}
\label{fig:band-eq}
\end{figure}
\par The van Hove singularity just above the Fermi level in the spin down channel is closely related
to the magnetic properties which we are going to discuss, so we explore the origin of it here. Because of
the similarity of the electronic structure in these four compounds only the bands and the
DOS of ZrFe$_2$ are shown here.
The zoomed-in band structure and the DOS of the minority states of ZrFe$_2$ are
shown in Figure \ref{fig:band-eq}. The peak in the minority states just above the Fermi level comes from
almost non-dispersive states. This energy range is highlighted by the rectangular block with height of
0.06 eV. The high DOS around the peak stems from the narrow bands in the directions along $W-K$
and $W-U$. This small dispersive states are stemmed from the specific atom arrangement of the compounds.
The Fe atoms in the C15 Laves phase form the so-called pyrochlore structure with
corner shared tetrahedrons. If we look at the net in the \{111\} layer, it is the Kagom\'{e} net with
alternatingly connected triangles and hexagons. As already shown by Johnston and Hoffmann \cite{johnston},
the high peaks in the DOS in a Kagom\'{e} net of iron atoms come from narrow bands with $d-\pi$ character. 
One band tight binding
calculation by Isoda \cite{isoda} discovers two non-dispersive degenerated states along the
$X-W$ line. Furthermore, there are two additional non-dispersive degenerated antibonding states along
all high symmetry directions. These results indicate that the spiking DOS is originated  from the
special geometrical arrangement of the Fe's.
\begin{figure}[!htbp]
\centering
\includegraphics[angle=-90,width=0.8\textwidth]{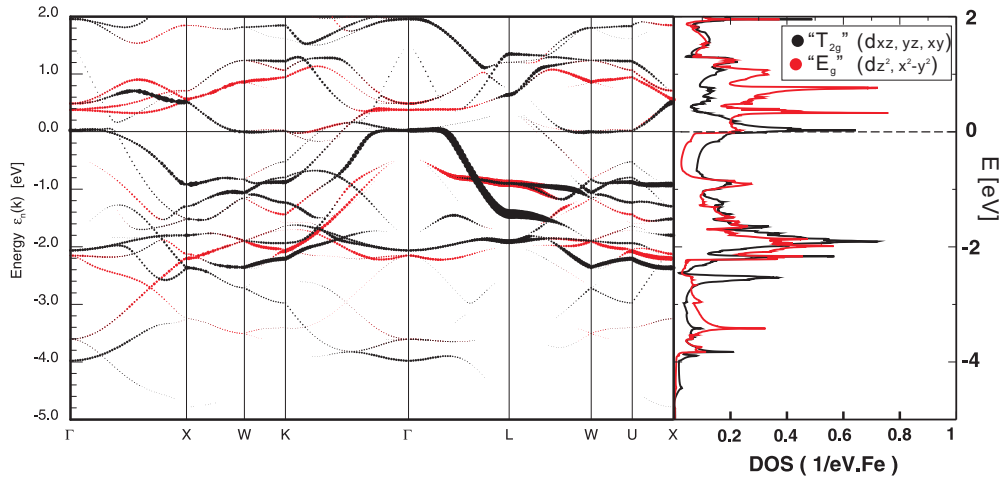}
\caption[The ``fat'' band and PDOS of the minority state of Fe in ZrFe$_2$.]{The ``fat'' band  and PDOS of the minority
d-state of Fe in ZrFe$_2$. The T$_{2g}$ states are from states with m$_l$=-2, -1 and 1,
while the E$_g$ states are from m$_l$=0 and 2.
}\label{fig:pdos-Fe}
\end{figure}
\par If we plot the m$_l$-resolved ``fat'' band (wighted contributions of atomic orbitals) and the partial DOS (PDOS)
of Fe as shown in Figure \ref{fig:pdos-Fe},
it is clear that at $\Gamma$-point the states with m$_l=0$ (d$_{z^2}$) and 2 (d$_{x^2-y^2}$) are degenerate and
m$_l=-2$ (d$_{xy}$), -1 (d$_{yz}$), and 1 (d$_{xz}$) are also degenerate. These two groups are denoted by E$_g$,
and T$_{2g}$, respectively.
It can be shown that the PDOS is divided into two catalogs:
One is from the three d orbitals
(m$_l=-2$, -1, and 1) and the other is from the rest two d orbitals (m$_l$=0, 2). It is shown that the strongest van 
Hove singularity
in the PDOS just above the Fermi
level of the minority spin state is solely from the T$_{2g}$ states. It is quite understandable because
the T$_{2g}$ states form quite strong $\pi$ binding states along
each atomic chain with nearest neighbor interactions.
The antibonding state of these $\pi$ orbitals is the source of the quite spiking feature of the DOS
in the cubic Laves phase compounds as analyzed by Johnston and Hoffman \cite{johnston}. From another point of view,
if we rotate the coordinates and put the z-direction to the diagonal direction of the cube, then the lattice can
be described by a trigonal lattice. There is a one dimensional irreducible
representation A$_{1g}$ with basis of d$_{z^{\prime 2}}$ for the point group of Fe site, of which is D$_{3d}$. It 
turns out that the d$_{z^{\prime 2}}$ orbital comes solely
from a linear combination of the original T$_{2g}$ orbitals. As discussed by Isoda \cite{isoda} by single orbital tight binding calculations,
which is naturally a one dimensional representation, the antibonding orbitals are non-dispersive in all high symmetric directions
of the BZ. This implies that the one dimensional representation with d$_{z^{\prime 2}}$ orbitals as its basis should give
also quite spiking feature in the DOS.
\section{The magnetic moment variations under pressure and its electronic characters}
Because of the differences of the \emph{A} atoms, we can
naturally expect some differences among these compounds. Firstly the
lattice constants of these materials are more or less determined by
the atomic volume of $A$. Taking the atomic volume, defined by
(atomic weight/mass density), of the elements: Y$=19.89$, Zr$=14.06$,
Lu$=17.78$ and Hf$=13.41$ (cm$^3$/mol), respectively, we can
see that the lattice constants in Table \ref{tab:exp} follow the
same tendency. 
\par
\begin{figure}
\centering
\includegraphics[width=0.8\textwidth]{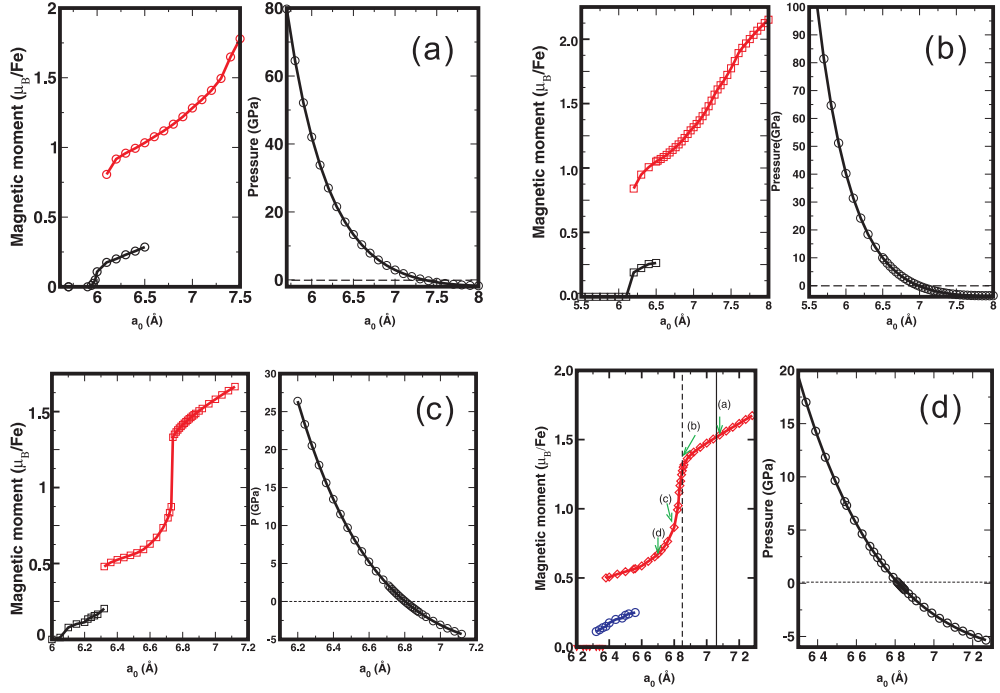}
\caption{\label{fig:mom-p}(Color online) magnetic moment variations of
AFe$_2$ (A = Y(a), Lu(b), Hf(c) and Zr(d)) at different lattice constant. The corresponding
hydrostatic pressures at the lattice constants are shown at the right of each subfigures.}
\end{figure}

 Secondly their magnetic moments have different behaviors under pressure.
The dependence of the magnetic moment on the lattice constants are
shown in Figure \ref{fig:mom-p}. The corresponding hydrostatic
pressures are shown on the upper abscissas.

Obviously all of them show a decrease of the magnetic moment with
the decrease of the lattice constant as expected from the itinerant
electron magnetism, but the Hf and Zr compounds show a more rapid
decrease of the moment at a lattice constant around $6.8$ \AA\ (in
the vicinity of the equilibrium lattice constant), while the other
two show a gradual decrease at this low pressure. At high pressure,
all four compounds show at least one first order transition to a
lower or zero spin state.
The differences are quite understandable by examining the
differences of the electron numbers of the compounds under the
assumption that the electronic structure is not so much influenced
by the difference of the $A$ atoms. YFe$_2$, ZrFe$_2$, HfFe$_2$, and
LuFe$_2$ show basically similar DOS as discussed before. 
The difference of the electron number shifts the Fermi level in
these systems. Zr(4$d^2$) and Hf(5$d^2$) have one more
\emph{d}-electron than Y(4$d^1$) and Lu(5$d^1$), so the Fermi levels
of the former are shifted towards higher energy, closer to
the pronounced peak of the minority spin DOS as shown in Figure
\ref{fig:dos-eq}. This accounts for the low pressure instability of the moment.
\par
\begin{figure}[!htbp]
\centering
\includegraphics[width=0.8\textwidth]{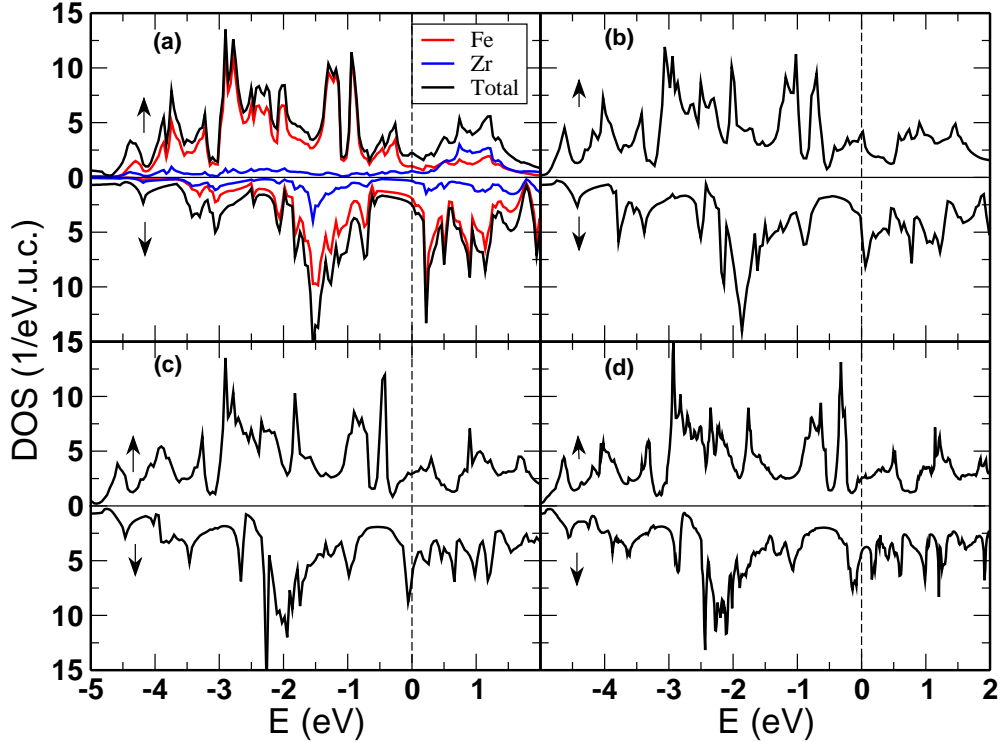}
\caption[DOS evolution of ZrFe$_2$ under pressure]{The total DOS and
partial DOS of ZrFe$_2$ at different lattice constants. From (a) to
(d), the lattice constants are 7.08, 6.85, 6.80, and 6.70 \AA,\
respectively. In (a), the partial DOS of Fe and Zr are also shown.
The Fermi level is indicated by the dashed vertical line at E=0
eV.}\label{fig:dos-zrfe2-a}
\end{figure}

 The four compounds show multi-step magnetic transitions. This process can be
understood by the particular DOS of these compounds. Taking ZrFe$_2$
as an example, the DOS at different lattice constants are shown in Figure
\ref{fig:dos-zrfe2-a} (a$\sim$d). The lattice constants of each
figure are indicated by the arrows in Figure \ref{fig:mom-p}(d)
with the corresponding labels of (a), (b), (c), and (d).

From Figure \ref{fig:dos-zrfe2-a}(a), it is obvious that at the
experimental lattice constant the DOS of the up spin, contributed
mainly from Fe, has a gradual increase below the Fermi level,
while the DOS of the down spin has a wide ($\sim$ 0.8 eV) dip below
and a sharp increase just above $E_F$.
Applying pressure will broaden the band and reduce the width of the
dip and decrease the magnetic moment. Thus the exchange splitting is
reduced. The DOS of the up spin and down spin moves towards each
other.
This gradual decrease of the magnetic moment is shown in Figure
\ref{fig:mom-p}(d) between the arrows (a) and (b). The gradual
decrease of the magnetic moment continues until
the Fermi level passes through the high DOS peak of the minority
spins, seeing Figure \ref{fig:dos-zrfe2-a}(b) and (c). Then the
magnetic moment is rapidly reduced, as shown in Figure
\ref{fig:mom-p}(d) when the lattice constant is between 6.85 (arrow
(b)) and 6.80 (arrow (c)) \AA. 

\par Very small magnetic moment is developed under the high pressures in
all of these compounds, as can be observed in Fig. \ref{fig:mom-p}.
This different behaviors when approaching the quantum phase
transition can be of interest both experimentally and theoretically.
For the second order phase transition, the effect of fluctuation was
shown to lead to novel electronic ground states in magnetic metals
such as magnetically mediated superconductivity, partial or
quadrupolar order and non-Fermi liquid phases. For the first order
phase transition, as summarized by Pfleiderer \cite{pfleiderer}, it
is interesting for a number of reasons: It can drive novel
electronic states, novel types of low lying excitations, or signal
the existence of subtle quantum correlation effects. In general,
peaks of DOS (van Hove singularities) near the Fermi level in all
real materials result in a ragged free energy landscape. The
shape of the DOS thus has a connection with the order of the quantum phase
transition. It is necessary to give some hints about the order of
the transition based on our calculations of electronic structures.
By the simplified Stoner model, the magnetic free energy of the
system in the rigid band model is expressed by \cite{shimizu65}
\begin{equation}
E(m)=\int_0^m\Delta\xi(m')dm'-\frac{1}{4}Im^2,
\end{equation}
where $\Delta\xi(m)$ is exchange splitting as a function of magnetic
moments $m$, and $I$ is the Stoner parameter. The $\Delta\xi(m)$ can
be expanded as a power series of $m$:
\begin{equation}
\Delta\xi(m)=a_1 m+a_3 m^3+a_5m^5\cdots
\end{equation}
where
\begin{eqnarray}
a_1&=&\frac{1}{2}\bar N_1^{-1}\label{eq:a1}\\
a_3&=&\frac{1}{3}(3\bar N_2^2\bar N_1^{-5}- \bar N_3\bar N_1^{-4})\label{eq:a3}\\
\nonumber a_5&=&\frac{2}{5!}(105\bar N_2^4\bar N_1^{-9}-105\bar N_3\bar N_2^2\bar N_1^{-8} \\
&&+10\bar N_3^2N_1^{-7}+15\bar N_4\bar N_2\bar N_1^{-7}-\bar N_5\bar
N_1^{-6}).\label{eq:a5}
\end{eqnarray}
$\bar{N}_i$ is defined as the $(i-1)$-th order derivative of the
density of states at the Fermi level with respect to the energy
\par Then the free energy is
\begin{equation}\label{eq:fm}
E(m)=\frac{1}{2}(a_1-\frac{I}{2})m^2+\frac{1}{4}a_3m^4+\frac{1}{6}a_5m^6\cdots.
\end{equation}
The stability of the phase can be discussed in line with Landau's
theory of second order phase transitions. Magnetic instability is
necessarily given by the condition that $a_1^\prime=
a_1-\frac{I}{2}\leq 0$, which is equivalent to the Stoner criterion
$IN(E_F)\geq
1$ by considering Equ. (\ref{eq:a1}). 
\par The necessary condition to have a first order transition is
$a_1-I>0$, $a_3<0$, and $a_5>0$ if higher order terms than $m^5$ are neglected in
Equation (\ref{eq:fm}). This means the DOS at the Fermi level should
be sufficiently small (the Stoner criterion is not fully satisfied)
and the curvature of the DOS at E$_F$ is positive and large, so that
$\bar{N}_3$ is positive and large enough to give negative $a_3$,
otherwise, if $\bar N_3<0$, $a_3$ is definitely positive. These
first two conditions require that the Fermi level is at a narrow
valley of the DOS.
\par
\begin{figure}[!hbt]
\centering
\includegraphics[width=0.8\textwidth]{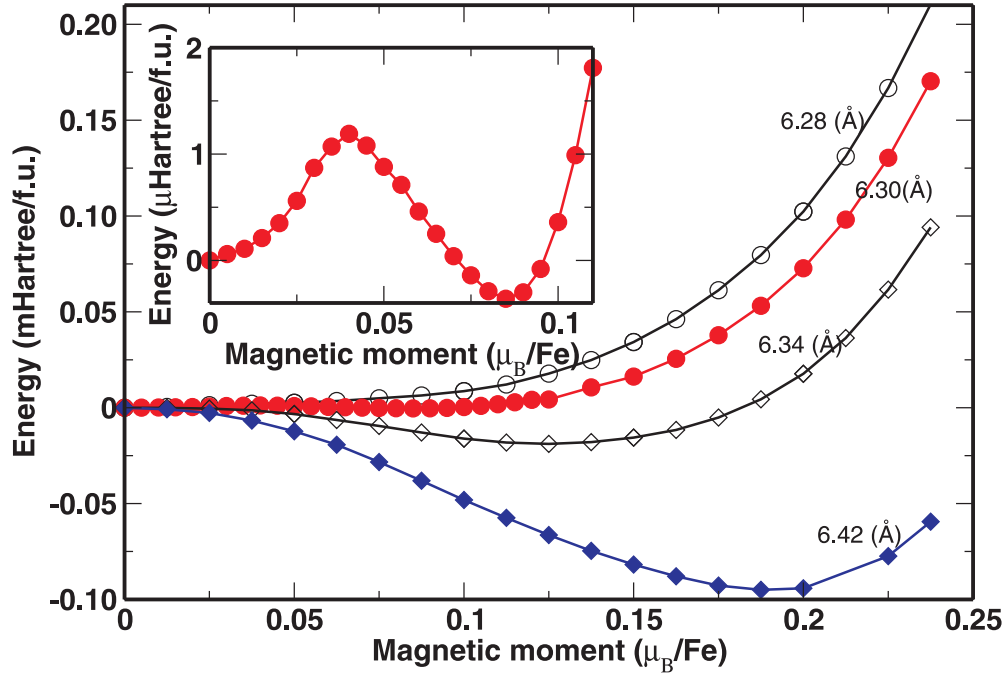}
\caption[The FSM energies of ZrFe$_2$ at the lattice constant around
6.30 \AA.]{The FSM curves of ZrFe$_2$ at the lattice constants
around 6.30 \AA. The inset shows the enlarged curve at the lattice
constant a=6.30 \AA. It clearly shows that magnetic and nonmagnetic
solutions coexist at this lattice constant. The data in this figure
are obtained with 3107 k-points in the IBZ.}\label{fig:fsm-zrfe2}
\end{figure}

\begin{figure}[!hbtp]
\centering
\includegraphics[width=0.8\textwidth]{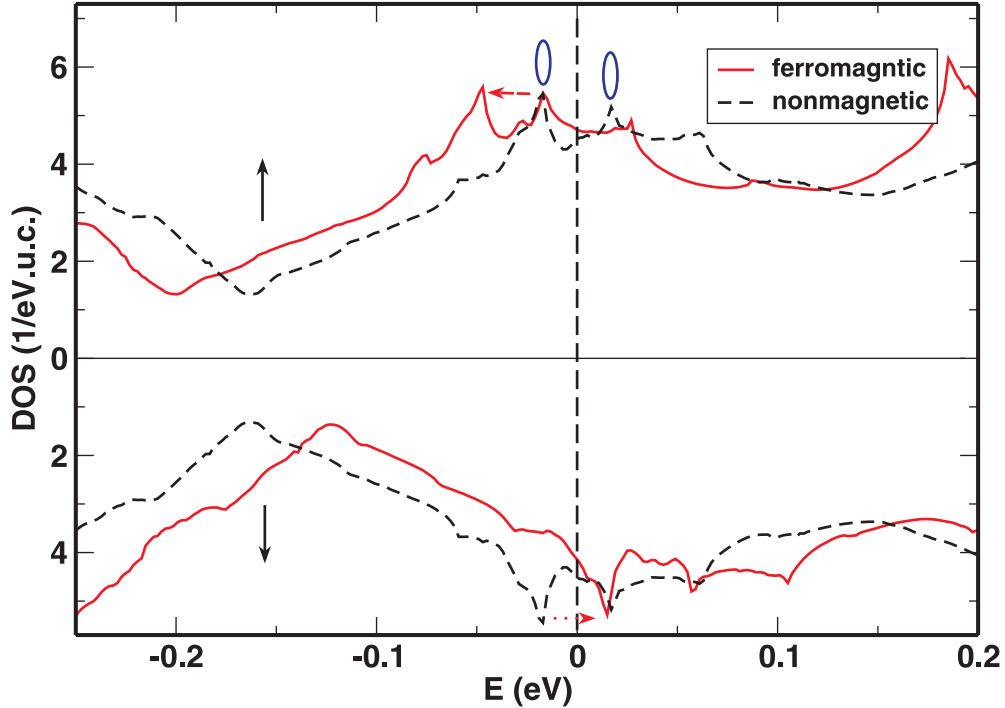}
\caption[The DOS of nonmagnetic state  and ferromagnetic state at
a=6.30 \AA.]{The DOS of nonmagnetic state (dashed lines) and
ferromagnetic state (red lines) of ZrFe$_2$ at a=6.30 \AA. The
horizontal dashed arrows show the relative shift of the DOS of the
up and down spin subbands. The two ellipses indicate the two peaks
around the Fermi level which cause the first order magnetic
transition.}\label{fig:dos-zrfe2m0}
\end{figure}

 Direct FSM calculation results and the corresponding DOS to analyze
the transition were added to the above qualitative analysis. The first example is ZrFe$_2$, which shows the first
order transition to the non-magnetic state. The FSM energy curves
are shown in Figure \ref{fig:fsm-zrfe2} at lattice constants around
the transition point.
The $E(m)$ curves at a=6.30 \AA\ are enlarged in the inset. It clearly
shows two energy minima at m=0 and m=0.085 $\mu_B$/Fe. The DOS of
the related nonmagnetic and magnetic solutions are shown in Figure
\ref{fig:dos-zrfe2m0}. It is clear that the Fermi level (the dashed
vertical line in the figure) is at a dip (between two peaks marked
by two ellipses) of the nonmagnetic DOS. At the magnetic solution,
the two subbands are shifted against each other as shown by the
dashed horizontal arrows.
\par
\begin{figure}[!hbt]
\centering
\includegraphics[width=0.75\textwidth]{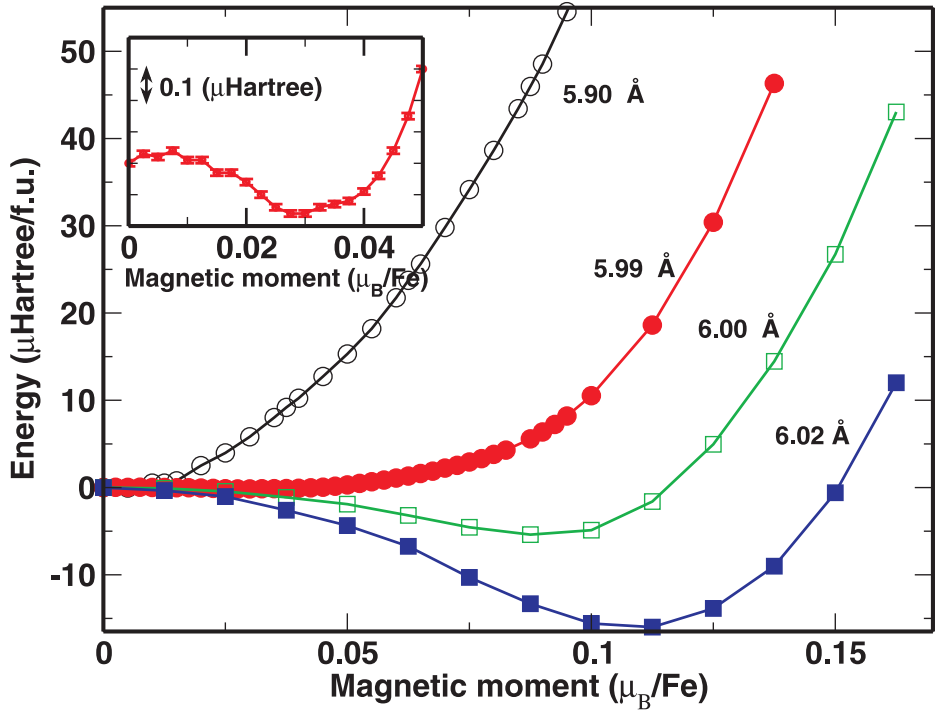}
\caption[The FSM energy of YFe$_2$ at lattice constant around 6.00
\AA. ]{The FSM energy of YFe$_2$ at lattice constants around a=6.00
\AA. The inset shows the zoomed-in curve at the lattice constant of
5.99 \AA\ with an error bar of 0.01 $\mu$Hartree.
}\label{fig:fsm-yfe2}
\end{figure}

\begin{figure}[!hbtp]
\centering
\includegraphics[width=0.8\textwidth]{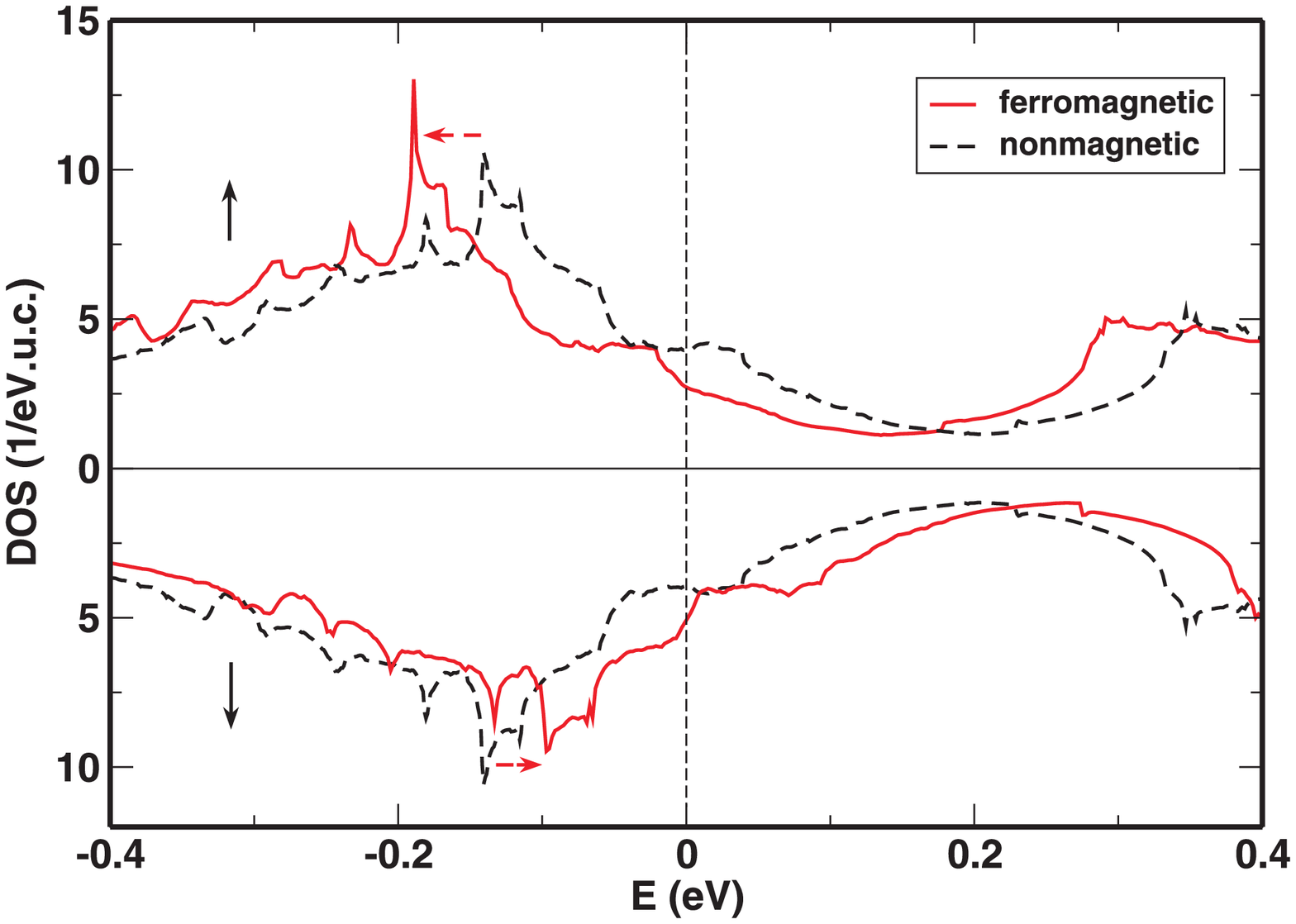}
\caption[The DOS of the nonmagnetic state and ferromagnetic state of
YFe$_2$ at a=5.98 \AA.]{The DOS of the nonmagnetic state (dashed
lines) and the ferromagnetic state (solid lines) of YFe$_2$ at
a=5.99 \AA. The horizontal dashed arrows show the relative shift of
the DOS of the up and down spin subbands. The Fermi level is shown
by the vertical dash.}\label{fig:dos-Yfe2m0}
\end{figure}

 The other example is YFe$_2$ where the magnetic transition is
the second order. The FSM curves are shown in Figure \ref{fig:fsm-yfe2}.
The energy minimum moves to zero when compressing the lattice as
shown in the figure. The energy curve at a=5.99 \AA\ is zoomed in
and shown in the inset. The FSM energy difference of small magnetic
moments reaches the accuracy limit guaranteed by the code. This is
the reason that we should use the DOS in order to discuss the
possible magnetic solutions. The DOS of nonmagnetic and
ferromagnetic states are shown in Figure \ref{fig:dos-Yfe2m0}.
It is clear that the ``valley'' character around the Fermi level is
missing compared with Figure \ref{fig:dos-zrfe2m0}. Rather, $E_F$ is
situated at a plateau which cannot have more than one magnetic
solutions. The other two compounds, LuFe$_2$ and HfFe$_2$ show 
similar second order transition.
\section{The magnetostriction Invar model}
Understanding the Invar effect, however, has been a problem for half of the century. More than
twenty different models have been published in the past 50 years for
explanation of the Invar effect. A general review about the
Invar effect can be found, for example, in handbooks edited by
Buschow and Wohlfarth \cite{wasserman}, and references therein.
One model called 2$\gamma$-model \cite{weiss} is based on the
hypothesis of Weiss that there exist two separated energy minima
with different volumes and magnetic states: High spin (HS) at high volume and
low spin (LS) at low volume states. First principle calculations of Fe$_3$Ni by Entel \cite{entel} and
other authors supported the 2$\gamma$-model. Entel argued that the
special position of the Fermi level in the minority band, being at
the crossover between nonbonding and antibonding states, is
responsible for the tendency of most Invar systems to undergo a
martensitic phase transition. Two minima binding curves should lead
to some discontinuity (a first order transition) in the pressure
dependence of certain physical properties, such as volume, magnetic
moment \emph{etc}., but this kind of discontinuity has never been
observed in Invar alloys as far as we know. This gives an obstacle in applying the
2$\gamma$-model to explain the Invar effect.
\par
\begin{figure}[!hbt]
\centering
\includegraphics[width=0.6\textwidth]{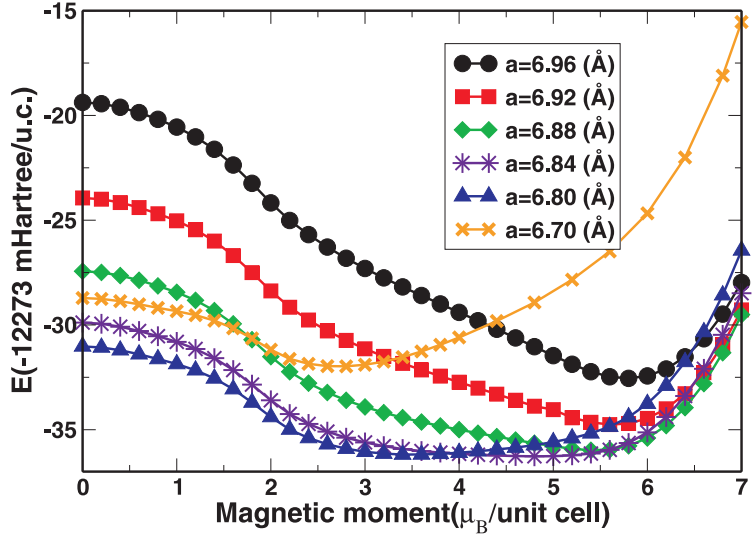}
\caption{The FSM energy curves of ZrFe$_2$ near the HS-LS transition
regions.} \label{fig:fsm2}
\end{figure}

 The HS-LS transition can also be continuous and it is in the
Invar alloy like ZrFe$_2$ and HfFe$_2$ as in Figure \ref{fig:mom-p},
This point can be clearly illustrated by our FSM calculations. In the FSM
energy curves, the energy minimum shifts to the lower magnetic
moments as the lattice constant is decreased as in Figure
\ref{fig:fsm2}. Here the FSM energy curves of ZrFe$_2$ is taken as
an example. The quite flat FSM energy curves, which means a large
spin susceptibility, near the transition region, because the
average DOS at the Fermi level is
large. The reciprocal susceptibility, $\chi^{-1}_M=E^{''}(M)$, is
given by \cite{kueblerbook}
\begin{equation}
\chi^{-1}_M=\mu^{-2}_B(2N_{eff}^{-1}-I),
\end{equation}
where $I$ is the Stoner parameter.
Thermal excitations cause loss of the magnetic moment leading to a
magnetic transition from the HS state to the LS state. Therefore,
increase of the temperature leads to gradual loss of the
spontaneous volume expansion associated with the ferromagnetic
state. This gradual process, contrary to the two states (HS and LS)
in some Invar alloy (e.g. Fe$_3$Ni), will not cause any
discontinuity in the pressure dependence of physical properties. In our compounds ZrFe$_2$ and HfFe$_2$ the gradual
decrease of the magnetic moment is the essential difference,
compared with the discontinuity present in a typical Invar system as
Fe$_3$Ni. 
\par As an important character of Invar alloy, the spontaneous volume magnetostriction is calculated by Equ.(\ref{eq:omegas}). 
The results are listed in Table \ref{tab:exp}, together with the
experimental data available \cite{shiga}. The theoretical values
agree with the experimental ones in the sense that they are at the
same order.
The overshooting of the spontaneous volume magnetostriction
($\omega_s$) can partly be from the non-vanishing local magnetic moment above the
transition temperature in the experiments, while in our model it is
in a Pauli paramagnetic state where the spin moment is zero. The
cure for this problem requires a more realistic treatment of the
paramagnetic phase. It has been shown that a noncollinear
\cite{Schilfgaarde} or a disordered local moment (DLM)
\cite{crisan, Khmelevskyi03} model gives a better agreement with the
experiments. Nevertheless, the results presented here show the major
characteristics of Invar alloy: Compared with the compounds where no
Invar anomaly is observed, the spontaneous volume magnetostriction $\omega_s$ is larger. In typical
Invar alloy, such as Ni$_{35}$Fe$_{65}$ and Fe$_{72}$Pt$_{28}$,
$\omega_s(10^{-3})=18$ and $14.4$ \cite{shiga}, respectively. At the same time, we see that the values of $\omega_s$ of
YFe$_2$ and LuFe$_2$ are also large. Why do they not show Invar
anomalies?
\par Take a simple ansatz of the temperature dependent volume of
a magnetic solid below its Curie temperature (T$_c$),
\begin{equation}
V(T)=V_0+\alpha(T-T_c)V_0+V^m(m(\tau)),
\end{equation}
where $V_0$ is the volume at $T_c$, $\alpha$ is the ``non-magnetic''
thermal expansion coefficient from phonon and electron
contributions, and $V^m(m(\tau))$ is the magnetic contribution to
the volume variation, with normalized magnetic moment $m=M/M_s$ and
at normalized temperature $\tau=T/T_c$. $M_s$ is the saturation
magnetization at T=0. The thermal expansion now reads,
\begin{align}
\frac{dV(T)}{V_0dT}&=\alpha+\frac{dV^m(m(\tau))}{V_0dT}\\
&=\alpha+\frac{M_sdV^m(m)}{T_cV_0dM}\frac{dm(\tau)}{d\tau}\\
&=\alpha+\frac{M_s}{T_cV_0dM/dV^m}\frac{dm(\tau)}{d\tau}.
\end{align}
In order to have a zero thermal expansion $\frac{dV(T)}{dT}=0$, we
require that
\begin{equation}
\frac{dM}{dV^m}=-\frac{M_s}{T_cV_0\alpha}\frac{dm(\tau)}{d\tau}.
\end{equation}
This shows that the $M(V^m)$ curve should follow the same behavior
as $m(\tau)$, scaled by a factor of $-\frac{M_s}{T_cV_0\alpha}$.
As we know that the temperature dependent magnetic phase transition
is of second order, so $m(\tau)$ is a continuous function of $\tau$.
Thus $M(V^m)$ should also be continuous. On the other hand, the
large decrease of the moment should take place near the equilibrium
volume at T$_c$ because our reference point is T$_c$.
In order to show the Invar anomaly, the rapid decrease of the
magnetic moment should be near the equilibrium lattice constant at
ambient conditions. This requirement excludes the Y, Lu compounds from
Invar alloy, where the decrease of the magnetic moment begins too far away from the equilibrium volume.
Doping of suitable atoms which shifts the Fermi level to the proper place can make this transition meet the requirement. 
\section{Conclusions}
To conclude, we studied the electronic structure and the magnetic
moment behaviors of four cubic Laves phase iron compounds. The
magnetic moment is found to decrease when the lattice constant is
decreased, and finally disappears. The way of the magnetic moment approaching zero can be 
continuously and discontinuously depending on the geometrical characters of the density of states.
It can be understood by the Landau's expansion of the magnetic free energy. Invar anomalies in these compounds
can be partly explained by the current work when the fast continuous magnetic moment decrease 
at the decrease of the lattice constant was properly considered. 
\begin{acknowledgments}
Discussions with M. Richter are greatly acknowledged. One of the authors, W.X. Zhang, thanks DAAD
for the financial support. Financial support from ``863''-projects (2015AA03130102) and Research Grant of Chinese Central Universities (ZYGX2013Z001) are acknowledged.
\end{acknowledgments}

\end{document}